\documentstyle[12pt,epsf,amssymb]{article}
\begin{document}

\tolerance=5000

\def\pp{{\, \mid \hskip -1.5mm =}}
\def\cL{{\cal L}}
\def\be{\begin{equation}}
\def\ee{\end{equation}}
\def\bea{\begin{eqnarray}}
\def\eea{\end{eqnarray}}
\def\tr{{\rm tr}\, }
\def\nn{\nonumber \\}
\def\e{{\rm e}}

\begin{titlepage}

\begin{center}
\Large
{\bf Modified gravity with negative and positive powers of the curvature: 
unification of the inflation and of the cosmic acceleration.}

\vfill

\normalsize

\large{ 
Shin'ichi Nojiri$^\spadesuit$\footnote{Electronic mail: nojiri@nda.ac.jp, 
snojiri@yukawa.kyoto-u.ac.jp} and 
Sergei D. Odintsov$^{\heartsuit\clubsuit}$\footnote{Electronic mail:
 odintsov@ieec.fcr.es Also at TSPU, Tomsk, Russia}}

\normalsize

\vfill

{\em $\spadesuit$ Department of Applied Physics, 
National Defence Academy, \\
Hashirimizu Yokosuka 239-8686, JAPAN}

\ 

{\em $\heartsuit$ Institut d'Estudis Espacials de Catalunya (IEEC), \\
Edifici Nexus, Gran Capit\`a 2-4, 08034 Barcelona, SPAIN}

\ 

{\em $\clubsuit$ Instituci\`o Catalana de Recerca i Estudis 
Avan\c{c}ats (ICREA), Barcelona, SPAIN}

\end{center}

\vfill

\baselineskip=24pt
\begin{abstract}

\noindent 
The modified gravity, which eliminates the need for dark
energy and which seems to be stable, is considered. The terms with
positive
powers of the curvature 
support the inflationary epoch while the terms with negative powers of the
curvature serve as effective dark energy, supporting current cosmic
acceleration.  The
equivalent scalar-tensor gravity may be compatible with the simplest solar
system
experiments.

\end{abstract}

%\pacs{98.80.Hw,04.50.+h,11.10.Kk,11.10.Wx}


\end{titlepage}

\section{Introduction}

It grows the evidence that the universe is undergoing a phase of the
accelerated expansion at present epoch. 
The indications to cosmic acceleration appeared not only from the 
high redshift surveys of type Ia supernovae \cite{R} but also 
from the anisotropy power spectrum of the Cosmic Microwave Background
(CMB) \cite{B}. The favored explanation for this behavior 
is that the universe is presently dominated by some form of dark energy.
However, neither of existing dark energy models is completely
satisfactory. Moreover it is very hard to construct the theoretical basis 
for the origin of this exotic matter, which is seen precisely at current
epoch 
when one needs the source for the cosmic acceleration.

In the recent papers \cite{CDTT,CCS} (see also \cite{v,v1}) it was
suggested
the gravitational
alternative for the dark energy. It goes as following: the standard
Einstein action is modified at low curvature by the terms which 
dominate precisely at low curvature. The simplest possibility of this sort
is $1/R$ term, other negative powers of the curvature may be introduced as
well. Moreover, more complicated terms may be suggested like $1/(\Box
R+ \mbox{constant})$ etc. The only condition is to dominate over $R$ and to
produce the  cosmic acceleration consistent with current
astrophysical data. The interesting feature of the theories with negative
powers of curvature is that they may be expected from some time-dependent 
compactification of string/M-theory as it was demonstrated in \cite{sn}.
%%%
Moreover, quantum fluctuations in nearly flat spacetime may induce such terms 
in the same way as the  expansion of the effective action at large curvature 
predicts the terms with positive power of (non-local combinations 
of) curvature invariants (for a 
recent review, see \cite{Vass}). 

Clearly, having the gravitational foundation for description of current 
cosmic acceleration seems to be much more natural than the introduction 
by hands of the mysterious dark energy, cosmic fluid with negative
pressure. However, as any other theory pretending to describe current FRW 
universe such a modified gravitational theory should pass number of
consistency checks to be considered as the realistic theory. For instance, 
as any other higher derivative theory, the theory with $1/R$ may develop 
the instabilities \cite{Dolgov} (see, however, \cite{meng}
and discussion in sixth section). As the cosmic
acceleration from such
theory is also instable \cite{CDTT} the question appears: which of the
instabilities is more realistic or some more modifications of the theory
may be required? From another point, the gravitational action which is the
function of only curvature may be presented in a number of ways in
the equivalent form as scalar-tensor theory with one (or several) 
scalar(s). Then fundamental question  is: which action
corresponds to our physical world? For instance, it has been mentioned
\cite{chiba} that for the model \cite{CDTT} the equivalent BD action is
not acceptable as it is ruled out by the solar system experiments.
This indicates against the consideration of the initial modified gravity 
with $1/R$ term as physical theory.
In such circumstances it may be a good idea to try to search for other
variants of modified gravitational theory which still excludes the
need for the dark energy. Moreover, one can try to construct the theory 
which predicts the inflation at very early Universe and cosmic
acceleration currently in the combined set-up.

In the present paper we suggest the new model of modified gravity 
which contains  positive   and negatives powers of the curvature.
Symbolically, the lagrangian looks like ${\cal L}=R+R^m +1/R^n$ where $n$, $m$
 are positive 
(not necessary integer ) numbers.
The theory may be presented in the equivalent form as some scalar-tensor
gravity. At large curvature, the terms of the 
sort $R^m$
dominate. If $ 1<m<2$ the power law inflation occurs at early times.
If $m=2$ the anomaly driven (Starobinsky) inflation occurs at early times.
At intermediate region, the theory is Einstein gravity. Currently,
at low curvature the terms of the sort $1/R^n$ dominate.
These terms serve as gravitational alternative to dark energy 
and produce the cosmic acceleration. It is remarkable that such modified
gravity does not suffer from the instabilities pointed out for 
the version with the lagrangian $L=R+1/R$. Moreover, it may pass the solar
system tests for scalar-tensor gravity.  

The paper is organized as follows. In the next section we discuss various 
forms of the action for modified gravity, both in Jordan and in Einstein
frames. It is shown also that such theory may have the origin in the
braneworld scenario by the fine-tuning of surface counterterms.
Section three is devoted to the study of simplest deSitter solutions 
for modified gravity. The occurrence of two deSitter phases is mentioned.
The properties of the scalar potential in the scalar-tensor formulation of
theory where scalar should be identified
with the curvature are investigated. In section four, FRW cosmology for
such a model (with or without matter) is discussed. It is shown that the
model naturally admits the unification 
of the inflation at early times and of the cosmic acceleration at late
times. In section six we demonstrate that higher derivative terms make 
the dangerous instabilities of original $1/R$ theory to become
much less essential 
at cosmological scales. Moreover, the model easily passes solar system 
constraints to scalar-tensor gravity. Some summary and outlook is
presented at final section.

\section{The  actions for the modified gravitational
theory}

Let us start from the rather general 4-dimensional action:
\be
\label{RR1}
S={1 \over \kappa^2}\int d^4 x \sqrt{-g} f(R)\ .
\ee
Here $R$ is the scalar curvature and $f(R)$ is some arbitrary function. 
Introducing the auxiliary fields $A$ and $B$, one may rewrite the
action 
(\ref{RR1}) as following:
\be
\label{RR2}
S={1 \over \kappa^2}\int d^4 x \sqrt{-g} \left\{B\left(R-A\right) + f(A)\right\}\ .
\ee
By the variation over $B$,  $A=R$ follows. Substituting it into
(\ref{RR2}), 
the action (\ref{RR1}) can be reproduced. Making the variation with
 respect to $A$ first, we obtain
\be
\label{RR3}
B=f'(A)\ ,
\ee
which may be solved with respect to $A$ as 
\be
\label{RR4}
A=g(B)\ .
\ee
Eliminating $A$ in (\ref{RR2}) with help of (\ref{RR4}), we obtain
\be
\label{RR5}
S={1 \over \kappa^2}\int d^4 x \sqrt{-g} \left\{B\left(R-g(B)\right) 
+ f\left(g(B)\right)\right\}\ .
\ee
Instead of $A$, one may eliminate $B$ by using (\ref{RR3}) and obtain
\be
\label{RR6}
S={1 \over \kappa^2}\int d^4 x \sqrt{-g} \left\{f'(A)\left(R-A\right) + f(A)\right\}\ .
\ee
At least classically, the two actions (\ref{RR5}) and (\ref{RR6}) are equivalent 
to each other. 

The action (\ref{RR5}) or (\ref{RR6}) may be called  the Jordan
frame action with auxiliary fields.
More convenient Einstein frame theory may be worked out as well. 
Under the conformal transformation
\be
\label{RR7}
g_{\mu\nu}\to \e^\sigma g_{\mu\nu}\ ,
\ee
$d$-dimensional scalar curvature is transformed as
\be
\label{RR8}
R^{(d)}\to \e^{-\sigma}\left(R^{(d)} - (d-1)\Box \sigma - {(d-1)(d-2) \over 4}
g^{\mu\nu}\partial_\mu \sigma \partial_\nu \sigma\right)\ .
\ee
Then for $d=4$, by choosing 
\be
\label{RR8b}
\sigma = -\ln f'(A)\ ,
\ee
the action (\ref{RR6}) is rewritten as
\be
\label{RR9}
S_E={1 \over \kappa^2}\int d^4 x \sqrt{-g} \left\{ R - {3 \over 2}\left({f''(A) 
\over f'(A)}\right)^2 
g^{\rho\sigma}\partial_\rho A \partial_\sigma A - {A \over f'(A)} 
+ {f(A) \over f'(A)^2}\right\}\ .
\ee
Using $\sigma=-\ln f'(A) = - \ln B$, the Einstein frame action looks like
\be
\label{RR10}
S_E={1 \over \kappa^2}\int d^4 x \sqrt{-g} \left( R - {3 \over 2}g^{\rho\sigma}
\partial_\rho \sigma \partial_\sigma \sigma - V(\sigma)\right)\ .
\ee
Here
\be
\label{RR11}
V(\sigma)= \e^\sigma g\left(\e^{-\sigma}\right) - \e^{2\sigma} f\left(g\left(\e^{-\sigma}
\right)\right)= {A \over f'(A)} - {f(A) \over f'(A)^2}\ .
\ee
This is the standard form of the scalar-tensor theories where scalar field
is fictitious one.
Although $S_E=S$, we denote the action given in the Einstein frame by $S_E$. 
In (\ref{RR8b}), it is assumed $f'(A)>0$.
Even if $f'(A)<0$,  $\sigma=-\ln \left| f'(A) \right|$ may be defined.
Then the sign in front of the scalar curvature becomes negative. In other words, 
anti-gravity could be generated.  

As a specific choice, we consider
\be
\label{RR12}
f(R)= R - {a \over \left( R - \Lambda_1 \right)^n} + b \left( R - \Lambda_2\right)^m \ .
\ee
Here we assume the coefficients $n,m,a,b>0$ but $n,m$ may be fractional.
Two last terms may be changed by the sum which
includes the terms with various negative and positive powers of the
curvature:
\be
\label{RR12BB}
f(R)= \sum_n a_n R^n \ .
\ee
Here $n$ can run from negative to positive values. In general,
$n$ needs not to be necessary integer. It may be also (positive or negative) 
irrational number, that is, $F(R)$ can be rather arbitrary function. 

When first term is absent, $n=m>1$, $-a=b$ and $\Lambda_1=\Lambda_2$ 
the evident duality symmetry appears. 
At very high curvature and supposing both cosmological constants to be
small, the higher derivative terms dominate.
Subsequently, with the decrease of the curvature the Einstein gravity 
dominates. For low curvature (depending on the choice of $\Lambda_1$) 
the negative power of the curvature may become the leading contribution to
the theory.

If $A$ or $R$ is large 
\be
\label{RR13}
\e^{-\sigma}=B\sim bm \left(A-\Lambda_2\right)^{m-1}\ .
\ee
Then if $m>1$ and $\e^{-\sigma}\to +\infty$ 
\be
\label{RR14}
V(\sigma)\sim - \left(bm\right)^{-{1 \over m-1}}\left(1 + {1 \over m}\right)
\e^{{m-2 \over m-1}\sigma} \ .
\ee
On the other hand when $A-\Lambda_1$ or $R-\Lambda_1$ is small, we obtain
\be
\label{RR15}
\e^{-\sigma}\sim an \left(A-\Lambda_1\right)^{-n-1}\ .
\ee
When $\e^{-\sigma}\to +\infty$ 
\be
\label{RR16}
V(\sigma)\sim - \left(an\right)^{{1 \over n+1}}\left(1 - {1 \over n}\right)
\e^{{n+2 \over n+1}\sigma} \ .
\ee
The above arguments indicate that $V(\sigma)$ is not a single-valued
function of $\sigma$ 
but at least, there are two branches for $m>2$.   
The above modified gravitational action will be our starting point 
in the attempt to construct the Universe where both phases: early time 
inflation and late time cosmic acceleration occur.

\section{The properties of the scalar potential}

 Let us discuss the properties of the scalar potential for 
$F(R)$  given by (\ref{RR12}). In this section, we concentrate on the 
case that the matter contribution can be neglected. 

With no matter and for the Ricci tensor $R_{\mu\nu}$ being
covariantly constant, the equation of motion corresponding to the action 
(\ref{RR1}) is:
\be
\label{R19}
0=2f(R) - Rf'(R)\ ,
\ee
which is the algebraic equation with respect to $R$. 
For the action (\ref{RR12})
\be
\label{RR17}
0=-R + {(n+2)a \over \left(R-\Lambda_1\right)^n} 
+ (m-2)b\left(R-\Lambda_2\right)^m\ .
\ee
Especially when $n=1$ and $m=2$, one gets
\be
\label{RR18}
R=R_\pm = {\Lambda_1 \pm \sqrt{\Lambda_1^2 +12a} \over 2}\ .
\ee
If $a>0$, one solution corresponds to deSitter space and another 
to anti-deSitter. 
If $-{\Lambda_1^2 \over 12}<a<0$ and $\Lambda_1>0$, both of solutions express the 
deSitter space. 
We may consider other cases, like $n=1$ and $m=3$. For simplicity, it is
chosen
$\Lambda_1=\Lambda_2=0$.  Then Eq.(\ref{RR17}) becomes
\be
\label{RR18b}
0=-R + {3a \over R} + bR^3\ .
\ee
For positive $a$ and $b$, Eq.(\ref{RR18b}) has two solutions in general.
If we further assume $a$ and $b$ is small, there is a  solution with large 
$R_l$ and the
 one with small $R_s$: 
\be
\label{RR18c}
R_l \sim b^{-{1 \over 4}}\ ,\quad R_s \sim \sqrt{3a}\ .
\ee
Since the square root of the scalar curvature 
corresponds to the expansion rate of the deSitter universe, the inflation
may be generated 
by a solution $R_l$ and the present cosmic speed-up may be generated by
$R_s$. One arrives at very interesting picture of the universe evolution 
where modification of gravity at high and low curvatures predicts both
(early and late time) phases of the accelerated expansion.

We now investigate the potential (\ref{RR11}) for the action 
(\ref{RR12}):
\be
\label{RR19}
V(A) = {{a\left\{(n+1)A - \Lambda_1\right\} \over \left(A-\Lambda_1\right)^{n+1}} 
+ b\left\{(m-1)A + \Lambda_2\right\}\left(A-\Lambda_2\right)^{m-1} \over
\left\{ 1 + {an \over \left(A-\Lambda_1\right)^{n+1}} + bm \left(A-\Lambda_2\right)^{m-1}
\right\}^2}\ .
\ee
As it is clear from Eq.(\ref{RR2}) that $A=R$, if we regard the scalar curvature $R$ 
in the action (\ref{RR1}) with the physical one, $A$ is nothing but the physical curvature 
itself (after equation of motion for $A$ is satisfied). 

When $A\to\pm \infty$
\be
\label{RR20}
V(A)\to {m-1\over bm^2}A^{-m+2}\ .
\ee
Therefore if $m<2$, $V(A)\to \infty$, if $m=2$, $V(A)\to {1 \over 4b}$, and 
if $m>0$, $V(A)\to 0$. On the other hand, when $A\to \Lambda_1$, one finds
\be
\label{RR21}
V(A)\to {\Lambda_1 \over an}\left(A-\Lambda_1\right)^{n+1} \to 0\ .
\ee

Since the potential (\ref{RR19}) is rather complicated, in the following, we only 
consider the case that $n=1$, $m=2$, $\Lambda_1=\Lambda_2=0$:
\be
\label{RR22}
V(A)={{2a \over A} + bA^2 \over \left(1 + {a \over A^2} + 2bA\right)^2}\ .
\ee
When $A\to\pm\infty$,  $V(A)\to {1 \over 4b}$. 
 $V(A)$ vanishes at
\be
\label{RR23}
A=0\ ,\quad -\left({2a \over b}\right)^{1 \over 3}\ .
\ee
The denominator of the potential (\ref{RR22}) vanishes at 
\bea
\label{RR24}
A&=& A_0\equiv - {1 \over 6b} + \alpha_+ + \alpha_- \ , \\
\alpha_\pm^3 &\equiv& {1 \over 2}\left\{ - {1 \over 54b^3} - {a \over 2b} 
\pm {1 \over 2b}\sqrt{\left(a + {1 \over 18b^2}\right)
\left(a+ {1 \over 54b^2}\right)}\right\}\nonumber \ .
\eea
When $A\to A_0$, the potential diverges. We should note $A_0<0$ if
$a,b<0$. Since 
\be
\label{RR24b}
V'(A)={2 \left( - {a \over A^2} + bA\right)\left(1 - {3a \over A^2}\right) 
\over \left(1 + {a \over A^2} + 2bA\right)^3}\ ,
\ee
 $V(A)$ has stationary points, where $V'(A)=0$, at
\be
\label{RR25}
A=\left({a \over b}\right)^{1 \over 3}\ ,\quad \pm\sqrt{3a}\ .
\ee
Then by summarizing the above analysis, we find the rough shape of the 
potential: The asymptotic value of the potential is ${1 \over 4b}$. The potential 
vanishes twice when $A$ vanishes and $A$ takes a negative value $A=
-\left({2a \over b}\right)^{1 \over 3}$ in (\ref{RR23}). 
The potential is singular when $A$ takes a negative value $A=A_0$ in 
(\ref{RR24}). The potential has three extrema. Two of them are given by positive 
$A$: $A=\left({a \over b}\right)^{1 \over 3}$, $\sqrt{3a}$ and 
one of the three extrema is given by negative $A$: $A=-\sqrt{3a}$. 

When $A$ is small, (\ref{RR8b}) shows that 
\be
\label{RR26}
\e^{-\sigma}={a \over A^2} + 2bA \sim {a \over A^2}\ ,
\ee
that is 
\be
\label{RR27}
A=\sqrt{a}\e^{{1 \over 2}\sigma}\ .
\ee
Since $A\to 0$ corresponds to $\sigma\to -\infty$ one may only consider
the region 
where $A$ and also physical scalar curvature are positive. Then the singularity at 
$A=A_0$  (\ref{RR24}) does not appear. 

Since the potential (\ref{RR22}) has the following form when $A$ is small, 
\be
\label{RR28}
V(A)\sim {2 \over a^2}A^3 \sim {2 \over \sqrt{a}}\e^{{3 \over 2}\sigma}\ , 
\ee
in the region where $A$ is small, the universe evolves as the power law inflation, 
as we will see later (\ref{RR45}). Eq.(\ref{RR28}) shows that the
potential is 
increasing function for small $A$. 
Since the potential is almost flat but monotonically increasing for large $A$, 
the following cosmological scenario  can be considered: Let us assume that 
the universe starts from large but finite positive $A$.  Since $A$
corresponds
to the physical scalar curvature $R$, the universe expands rapidly 
because $\sqrt{R}$ 
is the expansion rate of the deSitter space. After that $A$ rolls 
down to the small value and the rate of the expansion becomes small. When $A$ reaches 
the small $A$ region, universe begins the power law expansion.

\begin{figure}%[htbp]
\begin{center}
\epsffile{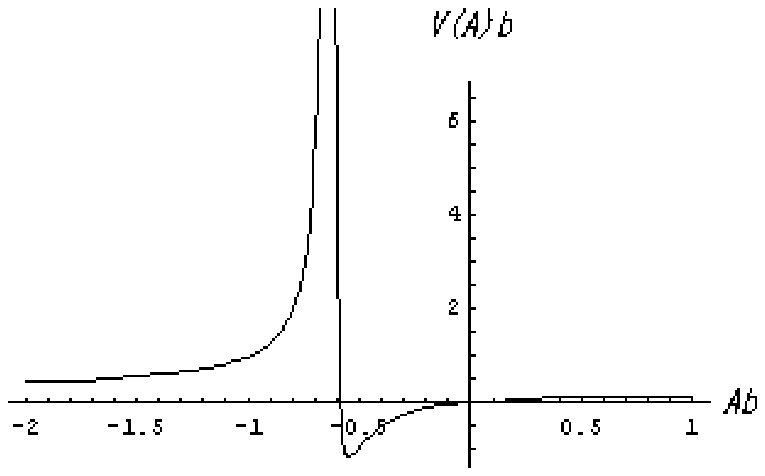}
\end{center}
\caption{\label{Fig1}
$V(A)b$ versus $Ab$ for $ab^2={1 \over 10}$. }
\end{figure}

\begin{figure}%[htbp]
\begin{center}
\epsffile{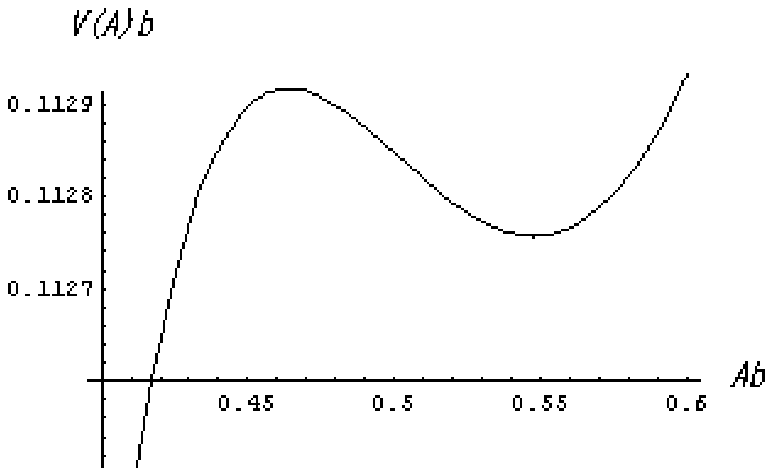}
\end{center}
\caption{\label{Fig2}
$V(A)b$ versus $Ab$ for $ab^2={1 \over 10}$. The behavior in the region 
$Ab\sim 0.5$.}
\end{figure}

\begin{figure}%[htbp]
\begin{center}
\epsffile{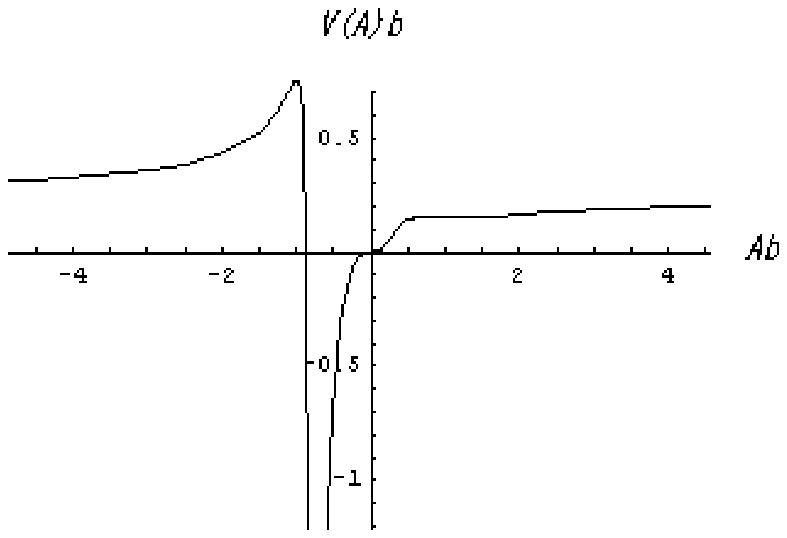}
\end{center}
\caption{\label{Fig3}
$V(A)b$ versus $Ab$ for $ab^2={1 \over 3}$. }
\end{figure}

We now give some typical shapes of $V(A)$. In Fig.\ref{Fig1}, 
the shape of the potential $V(A)b$ versus $Ab$ for $ab^2={1 \over 10}$ is given.  
When $A\to\pm\infty$, $V(A)$ approaches to the constant value ${1 \over
4b}$. 
As mentioned in (\ref{RR23}), $V(A)$ vanishes at $A=0$ and 
$A=-\left({2a \over b}\right)^{1 \over 3}<0$. 
When $A\to A_0<0$ in (\ref{RR4}), the potential diverges.
The negative $A$ minimum corresponds to $A=-\sqrt{3a}$. 
In Fig.(\ref{Fig2}),  the region $A\sim 0.5$ is considered. Then  
two extrema occur, which correspond to $A=\left({a \over b}\right)^{1
\over 3}$ and 
$A=\sqrt{3a}$ ( $\left({a \over b}\right)^{1 \over 3}
<\sqrt{3a}$). If the universe starts at $A=\sqrt{3a}$, which is locally
stable, the 
universe is in deSitter phase. If by some mechanism like the thermal
fluctuations etc., $A$ becomes smaller, 
the universe evolves to the power law inflation. The universe may start with 
$A=\left({a \over b}\right)^{1 \over 3}$, where the universe is unstable. 
We should note that the potential with $ab^2={1 \over 10}$ is bounded below. 
In Fig.\ref{Fig3}, the shape of the potential $V(A)b$ versus $Ab$ for $ab^2={1 \over 3}$ 
is given. The behavior when $A\geq 0$ is qualitatively not so changed
from 
$ab^2={1 \over 10}$ case. In the region $A<0$, however, the potential becomes unbounded 
below. When $ab^2={1 \over 10}$, the potential is bounded below 
since $A_0 < -\sqrt{3a}$. Note that $A=-\sqrt{3a}$ corresponds to the minimum. 
When $ab^2={1 \over 3}$, we have $A_0 < -\sqrt{3a}$ and $A=-\sqrt{3a}$ corresponds to 
the maximum, then the potential becomes unbounded below. Note once more
that the above results have physical meaning in terms of our modified
gravity with negative powers of curvature only when equation of motion for
the field $A$ itself is satisfied. (Only in this case $A$ may be identified
with the curvature of the physical metric).

Thus, we discussed the properties of the scalar potential corresponding to
quite simple version of the modified gravity. Similarly, the properties 
of other models of such sort may be analyzed.

\section{FRW cosmology in modified gravity}

Some simple properties of FRW cosmology in modified gravity may be
easily addressed.  One may add the matter to the action (\ref{RR1})
with the matter action  denoted by $S_{(m)}$. Then the energy-momentum
tensor 
$T_{(m)\mu\nu}$ can be defined by
\be
\label{RR29}
T_{(m)\mu\nu}=-{2 \over \sqrt{-g}}{\delta S_{(m)} \over \delta g^{\mu\nu}}\ .
\ee
After the rescaling the metric (\ref{RR7}), the energy-momentum tensor 
$T_{(mE)\mu\nu}$ in the Einstein frame (\ref{RR10}) is related with 
$T_{(m)\mu\nu}$ by
\be
\label{RR30}
T_{(mE)\mu\nu}=\e^{\sigma}T_{(m)\mu\nu}\ .
\ee
Defining the matter energy density $\rho_{(m)}$ and the pressure $p_{(m)}$
by
\be
\label{RR31}
T_{(m)00}=-\rho_{(m)} g_{00}\ ,\quad T_{(m)ij}=p_{(m)} g_{ij}\ ,
\ee
the corresponding quantities $\rho_{(mE)}$ and $p_{(mE)}$ in the Einstein frame 
(\ref{RR10}) are given by
\be
\label{RR32}
\rho_{(mE)}=\e^{2\sigma}\rho_{(m)}\ ,\quad p_{(mE)}=\e^{2\sigma}p_{(m)}\ .
\ee
We now assume the metric in the physical (Jordan) frame is given in the FRW form:
\be
\label{RR33}
ds^2 = - dt^2 + \hat a(t)^2 \sum_{i,j=1}^3 \hat g_{ij} dx^i dx^j\ .
\ee
Here $\hat g_{ij}$ is the metric of the Einstein manifold, which is defined by 
the Ricci tensor $\hat R_{ij}$ constructed from $\hat g_{ij}$ by 
$\hat R_{ij}=k \hat g_{ij}$ with a constant $k$. From the 
conservation of the energy-momentum tensor $\nabla^\mu T_{\mu\nu}=0$ one
gets 
\be
\label{RR34}
0=\dot \rho_{(m)} + 3H \left(\rho_{(m)} + p_{(m)}\right)\ .
\ee
Here
\be
\label{RR35}
H={\dot{\hat a} \over \hat a}\ ,
\ee
and it is assumed $\rho_{(m)}$, $p_{(m)}$ and also $\sigma$ does not
depend on the spatial 
coordinates but on the time coordinate. The matter is chosen to be the
perfect fluid, 
which satisfies
\be
\label{RR36}
p_{(m)} = w \rho_{(m)} \ .
\ee
 Eq.(\ref{RR34}) can be solved in the usual way
\be
\label{RR37}
\rho_{(m)} = C\hat a^{-3(1+w)}\ ,
\ee
with a constant of the integration $C$. In the Einstein frame, from 
(\ref{RR32}), we find
\be
\label{RR38}
\rho_{(mE)} = C\hat{a_E}^{-3(1+w)} \e^{-{3w - 1 \over 2}\sigma}\ .
\ee
Here $\hat{a_E}$ is the scale factor in the Einstein frame
\be
\label{RR39}
\hat{a_E}=\e^{-{\sigma \over 2}}\hat a\ .
\ee
 FRW equation in the Einstein frame has the following form:
\be
\label{RR40}
3 H_E^2 + {3k \over 2\hat{a_E}^2}={\kappa^2 \over 2}\left(\rho_{(\sigma E)} 
+ \rho_{(mE)}\right) \ .
\ee
Here
\be
\label{RR40b}
H_E\equiv {\dot{\hat{a_E}} \over \hat{a_E}}
\ee
and $\rho_{(\sigma E)}$ expresses the contribution from the $\sigma$ field. 
\be
\label{RR41}
\rho_{(\sigma E)}\equiv {1 \over \kappa^2}\left({3 \over 2}\dot \sigma^2 + V(\sigma)
\right)\ .
\ee
With the Einstein frame metric denoted as $g_{E\mu\nu}$, one obtains 
\be
\label{RR42}
{1 \over \sqrt{- g_E}}{\delta S_{(m)} \over \delta \sigma} 
= - {1 \over \sqrt{- g_E}}g^{\mu\nu}{\delta S_{(m)} \over \delta g^{\mu\nu}} 
= {1 \over 2}g^{\mu\nu}T_{\mu\nu}\sqrt{g \over g_E} \ .
\ee
Further using (\ref{RR31}), (\ref{RR36}), (\ref{RR37}), and (\ref{RR39}), 
we find
\be
\label{RR42b}
{1 \over \sqrt{- g_E}}{\delta S_{(m)} \over \delta \sigma} 
= {3w -1 \over 2}\rho_{(m)}\e^{2\sigma}
={3w -1 \over 2}C \hat{a_E}^{-3(1+w)}\e^{-{3w -1 \over 2}\sigma}\ .
\ee
Then the equation of motion for $\sigma$ in the Einstein frame, corresponding 
to the action $S_E + S_{(m)}$ ($S_E$ is given in (\ref{RR10})) has the following 
form: 
\be
\label{RR43}
0=3\left(\ddot \sigma + 3 H_E \dot \sigma\right) + V'(\sigma) 
 - {3w - 1 \over 2}\kappa^2C\hat{a_E}^{-3(1+w)}\e^{-{3w -1 \over 2}\sigma}\ .
\ee

The question now is to solve the system of equations (\ref{RR40}) and
(\ref{RR43}). 
Before going to consider the coupling with matter, we first consider the case of 
the vacuum where $C=0$. Especially when the potential is given by (\ref{RR28}) 
as $A=R$ is small, a solution is given by
\be
\label{RR44}
\hat{a_E}=\hat{a_{E0}}\left({t_E \over t_0}\right)^{4 \over 3}\ ,
\quad \sigma = - {4 \over 3}\ln {t_E \over t_0}\ .
\ee
Here $t_E$ is the time coordinate in the Euclidean frame, which is related with 
the time coordinate $t$ in the (physical) Jordan frame by $\e^{\sigma \over 2}dt_E 
= dt$. As a result
\be
\label{RR45b}
3t_E^{1 \over 3}=t\ ,
\ee
and even in the physical (Jordan) frame the power law inflation occurs
\be
\label{RR45}
\hat a = \e^{{\sigma \over 2}} \hat{a_E} \propto t_E^{2 \over 3}\propto t^2\ ,
\ee
which is consistent with the result in \cite{CDTT}.\footnote{
$\phi$ in \cite{CDTT} can be identified as $\sigma = - \sqrt{2 \over 3}{\phi 
\over M_p}$.} Hence, at small curvature the (instable) cosmic acceleration 
is predicted by the terms containing inverse curvature.
%%%%%%%%%%%%
If the present universe corresponds to the above power law inflation, the curvature 
of the present universe should be small compared with that of the deSitter universe 
solution in (\ref{RR18}) with $\Lambda_1=0$. As the Hubble constant in the 
present universe is $\left(10^{-33}{\rm eV}\right)^{-1}$, the parameter $a$, 
which corresponds to $\mu^4$ in \cite{CDTT}, should be much larger than 
$\left(10^{-33}{\rm eV}\right)^4$. 
%%%%%%%%%%%

One may also consider the case that the curvature is large. When $A$ or
the curvature is 
large, the potential (\ref{RR22}) becomes a constant: $V(A)\to {1 \over 4b}$. 
Then from (\ref{RR43}), we may assume, if we neglect the contribution from matter, that 
$\dot \phi$ is small. From (\ref{RR40}) it follows
\be
\label{RR46b}
3 H_E^2 + {3k \over 2\hat{a_E}^2}\sim {\kappa^2 \over 8b}\ ,
\ee
which shows that the spacetime is deSitter. For the case of $k=0$, $H_E$
becomes a 
constant:
\be
\label{RR46c}
H_E=\sqrt{\kappa^2 \over 24b}\ .
\ee
Then universe expands exponentially.

The matter contribution to the energy-momentum tensor may be accounted for 
in the same way. When it is dominant compared with one from $\sigma$, 
 Eqs.(\ref{RR40}) and (\ref{RR43}) can be reduced 
as 
\bea
\label{RR46}
0&=& 3 H_E^2 + {3k \over 2\hat{a_E}^2}-{\kappa^2 \over 2}
C\hat{a_E}^{-3(1+w)} \e^{-{3w - 1 \over 2}\sigma} \ ,\\
\label{RR47}
0&=&3\left(\ddot \sigma + 3 H_E \dot \sigma\right) 
 - {3w - 1 \over 2}\kappa^2 C\hat{a_E}^{-3(1+w)}\e^{-{3w -1 \over 2}\sigma}\ .
\eea
When $w={1 \over 3}$, which corresponds to the radiation, from (\ref{RR47}), 
one finds
\be
\label{RR48}
\dot\sigma = \dot\sigma_0\left({\hat{a_{E0}} \over \hat{a_E}}\right)^3\ .
\ee
Here $\dot\sigma_0$ and $\hat{a_{E0}}$ are constants. Eq.(\ref{RR48}) 
expresses the redshift of $\dot\sigma$. 
When $w=0$, which corresponds to the dust, if we can regard $\sigma$ is almost 
constant, we find
\be
\label{RR49}
\dot\sigma = \dot\sigma_0\left({\hat{a_{E0}} \over \hat{A_E}}\right)^3\ .
- {C\kappa^2\left(t-t_0\right) \over 6}\hat{a_E}^{-3(1+w)}\e^{-{3w -1 \over 2}\sigma}
\ee
The obtained results are not changed from those in \cite{CDTT} where the
possibility of cosmic acceleration in $1/R$ model was established. 

When $n=1$ or in general $n$ is an odd integer, as is clear 
from (\ref{RR16}), $A=0$ is not a minimum nor local minimum although 
$V'(A)=0$ at $A=0$. Of course, since $A\to 0$ corresponds to $\sigma\to + \infty$, 
it might be unnecessary to consider the region $A<0$. In the 
region $A\geq 0$, $A=0$ can give a minimum of the potential. 
On the other hand, when $n$ is an even integer,  $A=0$ is at least local 
minimum. Then if inflation occurs in the region $A\sim 0$, the inflation 
should be stable. 

We now consider the general case that $f(A)$ is given by (\ref{RR12}). 
When $R=A\sim \Lambda_1$ is small, the potential is given by (\ref{RR16}). 
Neglecting the contribution from the matter fields,  solving
(\ref{RR40}) and
(\ref{RR43}), we obtain, instead of (\ref{RR44}), 
\be
\label{RR44g}
\hat{a_E}=\hat{a_{E0}}\left({t_E \over t_0}\right)^{(n+1)(2n+1) \over (n+2)^2}\ ,
\quad \sigma = - {2(n+1) \over (n+2)}\ln {t_E \over t_0}\ .
\ee
Instead of (\ref{RR45b}) the physical time is 
\be
\label{RR45bg}
(n+2)t_E^{1 \over n+2}=t\ .
\ee
Then  the power law cosmic acceleration occurs in the
physical (Jordan) frame:
\be
\label{RR45g}
\hat a \propto t^{(n+1)(2n+1) \over n+2}\ .
\ee
It is quite remarkable that actually any negative power of the curvature
supports the cosmic acceleration. This gives the freedom in modification
of the model to achieve the consistency with  experimental tests of
newtonian gravity.

On the other hand, if the scalar curvature $R=A$ is large, the potential is given 
by (\ref{RR14}). Solving  (\ref{RR40}) and (\ref{RR43}) again, one obtains 
\be
\label{RR44gL}
\hat{a_E}=\hat{a_{E0}}\left({t_E \over t_0}\right)^{(m-1)(2m-1) \over (m-2)^2}\ ,
\quad \sigma = - {2(m-1) \over (m-2)}\ln {t_E \over t_0}\ ,\quad
(m-2)t_E^{-{1 \over m-2}}=t\ .
\ee
Thus in the physical (Jordan) frame, the universe shrinks with the power 
law if $m>2$\footnote{For $m=2$ the well-known anomaly driven
(Starobinsky) inflation occurs.}:
\be
\label{RR45gL}
\hat a \propto t^{-{(m-1)(2m-1) \over m-2}}\ .
\ee
Of course, if we change the arrow of time by $t\to t_0 - t$, the inflation occurs 
with the inverse power law and at $t=t_0$, the size of the universe
diverges. It is remarkable that when $m$ is fractional (or irrational) and $1<m<2$, 
Eqs.(\ref{RR44gL}) and (\ref{RR45gL}) are still valid. Then the power in Eq.(\ref{RR45gL}) 
becomes positive, the universe evolves with the (fractional) power law expansion.

Anyway, it is interesting that our model may unify both phases: early time
inflation (for $m<2$ or $m=2$) and current cosmic acceleration.

In \cite{CDTT},  from the dimensional analysis it is predicted for $\mu^4
= a$
($a$ for $n=1$ case) 
 to be $\mu \sim H_0 \sim 10^{-33}$ eV. Here $H_0$ is the Hubble constant in 
the present universe. From the dimensional analysis, one may
choose 
\be
\label{RR45gLb}
a=\left( 10^{-33}\, \mbox{eV}\right)^{2(n+1)}\ ,\quad  
b=\left( 10^{-33}\, \mbox{eV}\right)^{2(-m+1)}\ .
\ee
In principle, these parameters should not be small ones,
unless this is predicted by experimental constraints.

Adding the (dominant) matter contribution to the energy-momentum tensor, 
 the situation is not changed from $n=1$ and $m=1$ case and 
the case of \cite{CDTT}. We obtain (\ref{RR48}) and (\ref{RR49}) again.
Thus, the possibility of cosmic acceleration in the model with negative
powers of the curvature is demonstrated. 
Moreover, terms with positive powers of the curvature in such a model may 
realize the inflation at early times.

\section{Simplest tests for modified gravity}

In this section, we discuss the (in)stability of our higher
derivative model under the perturbations. Simplest constraint to
the theory parameters from the equivalent BD-type gravity
is also analyzed.
 
 In \cite{Dolgov}, small gravitational object  like 
the Earth or the Sun in the model \cite{CDTT} is considered. It has
been shown that the 
system quickly becomes instable. 

The general\footnote{We consider the case that the Ricci tensor is not covariantly constant} 
equation of motion corresponding the action (\ref{RR1}) with matter is given by
\be
\label{RR56}
{1 \over 2}g_{\mu\nu} f(R) - R_{\mu\nu} f'(R) - g_{\mu\nu} \Box f'(R) 
+ \nabla_\mu \nabla_\nu f'(R) = - {\kappa^2 \over 2}T_{(m)\mu\nu}\ .
\ee
Here $T_{(m)\mu\nu}$ is the energy-momentum tensor of the matter. By multiplying $g^{\mu\nu}$ 
to (\ref{RR56}), one arrives at\footnote{
The convention of the signature of the spacetime here is different from that in 
\cite{Dolgov}.}
\be
\label{RR57}
\Box R + {f^{(3)}(R) \over f^{(2)}(R)}\nabla_\rho R \nabla^\rho R 
+ {f'(R) R \over 3f^{(2)}(R)} - {2f(R) \over 3 f^{(2)}(R)} 
= {\kappa^2 \over 6f^{(2)}(R)}T\ .
\ee
Here $T=T_{(m)\rho}^{\ \rho}$. Let $f(R)$ is given by 
(\ref{RR12}). Then in case of the Einstein gravity, where $a=b=0$, the solution of 
Eq.(\ref{RR57}) is given by 
\be
\label{RR58}
R=R_0\equiv -{\kappa^2 \over 2}T\ .
\ee
The perturbation around the solution (\ref{RR58}) may be addressed 
\be
\label{RR59}
R=R_0 + R_1\ ,\quad \left(\left|R_1\right|\ll \left|R_0\right|\right)\ .
\ee
Then by linearizing (\ref{RR57}), we obtain 
\bea
\label{RR60}
0&=&\Box R_0 + {f^{(3)}(R_0) \over f^{(2)}(R_0)}\nabla_\rho R_0 \nabla^\rho R_0 
+ {f'(R_0) R_0 \over 3f^{(2)}(R_0)} - {2f(R_0) \over 3 f^{(2)}(R_0)} 
 - {R_0 \over 3f^{(2)}(R_0)} \nn
&& + \Box R_1 + 2{f^{(3)}(R_0) \over f^{(2)}(R_0)}\nabla_\rho R_0 \nabla^\rho R_1 
+ U(R_0) R_1\ .
\eea
Here
\bea
\label{RR61}
&& U(R_0)\equiv \left({f^{(4)}(R_0) \over f^{(2)}(R_0)} - {f^{(3)}(R_0)^2 
\over f^{(2)}(R_0)^2}\right) \nabla_\rho R_0 \nabla^\rho R_0 + {1 \over 3}R_0 \nn
&& - {f^{(1)}(R_0) f^{(3)}(R_0) R_0 \over 3 f^{(2)}(R_0)^2} 
 - {f^{(1)}(R_0) \over f^{(2)}(R_0)} + {2 f(R_0) f^{(3)}(R_0) \over 3 f^{(2)}(R_0)^2} 
 - {R_0 f^{(3)} \over f^{(2)}(R_0)^2} \ .
\eea
If $U(R_0)$ is negative, the perturbation $R_1$ grows up exponentially with time. 
The system becomes instable. In the following, we neglect the terms of the
sort  
$\nabla_\rho R_0$. In case of the model \cite{CDTT}, where 
$b=\Lambda_1=\Lambda_2=0$
\be
\label{RR62}
U(R_0) = - R_0 + {R_0^3 \over 6a}\ .
\ee
In order to describe the universe acceleration in the present epoch, we may have \cite{CDTT}
\be
\label{RR63}
\mu^{-1}\equiv a^{-{1 \over 4}}\sim 10^{18} \mbox{sec} \sim \left( 10^{-33} \mbox{eV} 
\right)^{-1}\ .
\ee
%%%%%%%%%%%
(As discussed after Eq.(\ref{RR45}), if the present universe corresponds to the power law 
inflation, $a$ can be much larger than $\left(10^{-33} {\rm eV}\right)^4$.)
%%%%%%%%%%
By using the estimations in \cite{Dolgov},
\be
\label{RR64}
{R_0^3 \over a}\sim \left(10^{-26} \mbox{sec}\right)^{-2} \left({\rho_m \over 
\mbox{g\,cm}^{-3}}\right)^3\ ,\quad R_0 \sim  \left(10^3 \mbox{sec}\right)^{-2} 
\left({\rho_m \over \mbox{g\,cm}^{-3}}\right)\ ,
\ee
we find the second term in (\ref{RR62}) dominates. Here $\rho_m$ is the mass density of the 
gravitating body. Since $R_0$ is negative from (\ref{RR58}), $U(R_0)$
becomes negative and the microscopic gravitational instability occurs.

First we consider the case that $n=1$, $m=2$, and $\Lambda_1=\Lambda_2=0$. 
For
\be
\label{RR65}
b\gg {a \over \left| R_0^3 \right|}\ ,
\ee
one gets
\be
\label{RR66}
U(R_0)\sim {R_0 \over 3}<0\ .
\ee
Then the instability shows up again. We should note, however, that since 
$\left|{R_0 \over 3}\right| \ll {R_0^3 \over 6a}$ from (\ref{RR64}),
 the (macroscopic) instability development takes quite long time. In fact,
the time for instability to occur is significally 
improved  (by the order of $10^{29}$). Clearly, other higher derivative
terms may significally improve the estimation as is shown below.  

Eqs.(\ref{RR63}), (\ref{RR64}), and (\ref{RR65}) indicate that 
$b^{-1}\ll \left(10^{11} {\rm eV}\right)^2$ if we assume $\rho_m\sim 1 \mbox{g\, cm}^{-3}$.
Then  ${\kappa^2 \over b} \ll \left(10^{-17}\right)^2$, which does not 
seem to be consistent with the  bounds obtained from the
observation \cite{Hwang} 
(see also \cite{Mijic} where HD inflation was discussed). The action of
model in \cite{Mijic, Hwang} contains the Einstein-Hilbert 
term and the $R^2$-term. 
There are two ways to overcome this bound.
First of all, one can include the contributions from other terms
containing 
some powers of the curvatures and/or the contribution from the matter fields and 
the quantum effects like conformal anomaly during the inflation.
For instance, account of conformal anomaly gives the way for trace anomaly
driven inflation.
 The bound
might be changed. 
From another side, as it follows from  the discussion after
Eq.(\ref{RR45}), if the present accelerating universe 
corresponds to the power law inflation, $a$ should be much larger than 
$\left(10^{-33} {\rm eV}\right)^4$. If we choose $a$ to be large enough in 
(\ref{RR62}) so that $a\gg R_0^2$, the first term in (\ref{RR62}), which is 
positive, dominates. Then the instability  \cite{Dolgov} does not 
exist from the very beginning.
%%%%%%%%

For $n=1$, $m=3$, and $\Lambda_1=\Lambda_2=0$ 
one can assume
\be
\label{RR67}
b\gg {a \over  R_0^4 }\ ,
\ee
and
\be
\label{RR68}
U(R_0)\sim -{5R_0 \over 18} + {1 \over 18 b R_0}\ .
\ee
Then if 
\be
\label{RR69}
b< {1 \over 5 R_0^2}\ ,
\ee
$U(R_0)$ is positive and the system becomes stable. From (\ref{RR64}), we
find 
${1 \over R_0^2} \gg {a \over  R_0^4 }$. Then the condition (\ref{RR69}) is compatible 
with the assumption (\ref{RR67}). Thus, the addition of the terms with the
positive powers of the curvature may salvage the modified gravity as less
instabilities occur or their development is macroscopic and other effects 
may prevent them.

It has been mentioned in ref.\cite{chiba} that $1/R$ model which is
equivalent to some scalar-tensor gravity is ruled out as realistic theory 
due to the constraints to such theories. Let us study if our theory 
may pass these constraints.
%%%%%%%%%%%%%%%%%%%%%%%%%%%%%%%%%%%%%%%
The coupling of the $\sigma$-field 
with matter is always of the same order with gravity \cite{fla}.
Then if $\sigma$ 
field has a small mass, the present model cannot be realistic. 
One may consider the case that the present universe corresponds to one of the 
solutions  (\ref{RR18}) with $\Lambda_1=0$, which corresponds to a minimum 
in (\ref{RR25}) of the potential $V(A)$: $R=A=\sqrt{3a}$. We now 
calculate the square of scalar mass, which is proportional to
$V''(\sigma)$. When $A=\sqrt{3a}$
\bea
\label{RF1}
\left.{d^2 V(\sigma) \over d\sigma^2}\right|_{A=\sqrt{3a}}&=&
\left\{\left({d\sigma \over dA}\right)^{-2}
\left.{d^2 V(A) \over dA^2}\right\}\right|_{A=\sqrt{3a}} \nn
&=&{ \sqrt{3a} \left({1 \over 3} + 2b\sqrt{3a}\right)^2 \over 
\left( - {1 \over 3} + b\sqrt{3a}\right)\left({4 \over 3} 
+ 2b\sqrt{3a}\right)^3} \ .
\eea
Choosing
\be
\label{SS5}
\beta \sim {1 \over 3 \sqrt{3a}}\ ,
\ee
the mass of $\sigma$ becomes large. Thus, HD term
may help to pass the solar system tests\cite{will}. 

%%%%%%%%%%%%%%%%%%

\section{Discussion}

In summary, we considered the modified gravity which naturally unifies two
expansion
phases of the universe: inflation at early times and cosmic acceleration
at current epoch. The higher derivatives terms with positive power of the
curvature are dominant at the early universe providing the inflationary
stage. The terms with negative power of the curvature serve as
gravitational alternative for the dark energy making possible the cosmic
speed-up. It is shown that such a theory is stable unlike to its more
simple cousin with only Einstein and $1/R$ terms. Moreover, it may pass
the simplest solar system constraint for scalar-tensor gravity equivalent 
to original modified gravity. Clearly, more checks of this theory should
be done in order to conclude if the model is realistic one or not. In
any case, there is some space for modifications by adding the terms 
with other powers of the curvature. Moreover, the use of other curvature
invariants (Ricci and Riemann tensors) may be considered as well.

It seems very attractive that modified gravity which may origin from
string/M-theory eliminates the need for dark energy. Clearly, deeper 
investigation in this extremely interesting direction connecting 
the cosmological constant problem, early time inflationary and current 
accelerating universe is necessary. It is a challenge to find the
solution for above fundamental cosmological problems from the first
principles. The search for true gravitational action may be the step in
this direction.

\section*{Acknowledgments} 

The research is supported in part by the Ministry of
Education, Science, Sports and Culture of Japan under the grant n.13135208
(S.N.), DGI/SGPI (Spain) project BFM2000-0810 (S.D.O.), RFBR grant 03-01-00105
(S.D.O.) and LRSS grant 1252.2003.2 (S.D.O.). 
 S.N. is indebted to all the members of IEEC, especially to
E. Elizalde, 
for the hospitality during the time when this work was finished. 

\appendix

\section{Modification of the gravity from the inflationary potentials}

It is very interesting that modified gravity may be often predicted 
by the standard inflationary cosmology with inflaton field if the
corresponding potential is not trivial.
Let us consider the model coupled with the scalar field with exponential 
potential:
\be
\label{RR50}
S=\int d^4 x \sqrt{-g} \left({M_P^2 \over 16\pi}R - {1 \over 2}g^{\mu\nu}
\partial_\mu \phi \partial_\nu \phi - V_0\e^{\sqrt{8\pi}\alpha \phi \over M_P}\right)\ .
\ee
Here $M_P$ is the 4-dimensional Planck mass.
It has been shown \cite{LM} that if $\alpha \leq \sqrt{2}$, the potential becomes 
shallow and supports the inflation. Recently there appeared some
arguments towards to possibility of the inflation and cosmic acceleration 
in the theory
based 
on the braneworld scenario with similar exponential potentials
\cite{braneS}. If 
the metric is rescaled
\be
\label{RR51}
g_{\mu\nu}\to \e^{{4 \over M_P}\sqrt{\pi \over 3}\phi}g_{\mu\nu}\ ,
\ee
the action (\ref{RR50}) can be rewritten as
\be
\label{RR52}
S=\int d^4 x \sqrt{-g} \left({M_P^2 \over 16\pi}\e^{{4 \over M_P}\sqrt{\pi \over 3}\phi}
R  - V_0\e^{\sqrt{8\pi}\left(\alpha + \sqrt{8 \over 3}\right){\phi \over M_P}}\right)\ .
\ee
Since the kinetic term for $\phi$ vanishes, we may regard $\phi$ to be an
auxiliary 
field. By the variation over $\phi$, one gets
\be
\label{RR53}
0={M_P \over 4\sqrt{3\pi}}\e^{{4 \over M_P}\sqrt{\pi \over 3}\phi}
R  - {V_0 \over M_P}\sqrt{8\pi}\left(\alpha + \sqrt{8 \over 3}\right)
\e^{\sqrt{8\pi}\left(\alpha + \sqrt{8 \over 3}\right){\phi \over M_P}}\ ,
\ee
which may be solved as 
\be
\label{RR54}
\e^{\sqrt{8\pi}\left(\alpha + \sqrt{2 \over 3}\right){\phi \over M_P}}
={M_P^2 \over 8\pi\sqrt{6}\left(\alpha + \sqrt{8 \over 3}\right)V_0} R\ .
\ee
By substituting $\phi$  (\ref{RR54}) into the action (\ref{RR52}), 
we obtain
\be
\label{RR55}
S={\sqrt{6}\left(\alpha + \sqrt{2 \over 3}\right)V_0 \over 2}\int d^4x \sqrt{-g}
\left\{{M_P^2 R \over 8\pi \sqrt{6}\left(\alpha + \sqrt{8 \over 3}\right)V_0}
\right)^{\alpha + \sqrt{8 \over 3} \over \alpha +\sqrt{2 \over 3}}\ .
\ee
Then the action (\ref{RR50}) is equivalent to the modified gravity with 
the fractional 
power ${\alpha + \sqrt{8 \over 3} \over \alpha + \sqrt{2 \over 3}}$ of 
the scalar curvature. If $-\sqrt{8 \over 3}<\alpha < -\sqrt{2 \over 3}$, 
the power of the curvature becomes negative again. 

Similarly, one can consider other inflationary potentials.
It is clear that many of them lead to modification of the gravitational
theory , supporting the point of view that inflaton may be not physical but
rather some fictitious scalar field.

\end{document}